\def\comp{{\rm C}\llap{\vrule height7.1pt width1pt depth-.4pt\phantom t}}
\def\Box{\kern1pt\vbox{\hrule height 1.2pt\hbox{\vrule width 1.2pt\hskip 3pt
   \vbox{\vskip 6pt}\hskip 3pt\vrule width 0.6pt}\hrule height 0.6pt}\kern1pt}
\def\gtwid{\mathrel{\raise.3ex\hbox{$>$\kern-.75em\lower1ex\hbox{$\sim$}}}}
\def\ltwid{\mathrel{\raise.3ex\hbox{$<$\kern-.75em\lower1ex\hbox{$\sim$}}}}
 
\documentstyle[epsfig,12pt]{article}

\begin{document}
\begin{titlepage}
\begin{flushright}
astro-ph/0109271 \\ UFIFT-HEP-00-21
\end{flushright}
\vspace{.4cm}
\begin{center}
\textbf{A Scalar Measure Of The Local Expansion Rate}
\end{center}
\begin{center}
L. R. Abramo$^{\dagger}$
\end{center}
\begin{center}
\textit{Theoretische Physik \\ Ludwig Maximilians Universit\"{a}t \\
Theresienstr. 37, \\ D-80333 M\"{u}nchen GERMANY}
\end{center}
\begin{center}
R. P. Woodard$^{\ddagger}$
\end{center}
\begin{center}
\textit{Department of Physics \\ University of Florida \\ 
Gainesville, FL 32611 USA}
\end{center}
\begin{center}
ABSTRACT
\end{center}
\hspace*{.5cm} We define a scalar measure of the local expansion rate based on
how astronomers determine the Hubble constant. Our observable is the inverse
conformal d'Alembertian acting on a unit ``standard candle.'' Because this
quantity is an integral over the past lightcone of the observation point it
provides a manifestly causal and covariant technique for averaging over small
fluctuations. For an exactly homogeneous and isotropic spacetime our scalar
gives minus one half times the inverse square of the Hubble parameter. Our
proposal is that it be assigned this meaning generally and that it be employed
to decide the issue of whether or not there is a significant quantum
gravitational back-reaction on inflation. Several techniques are discussed for
promoting the scalar to a full invariant by giving a geometrical description
for the point of observation. We work out an explicit formalism for evaluating
the invariant in perturbation theory. The results for two simple models are
presented in subsequent papers.
\begin{flushleft}
\end{flushleft}
\vspace{.4cm}
\begin{flushleft}
$^{\dagger}$ e-mail: abramo@theorie.physik.uni-muenchen.de \\
$^{\ddagger}$ e-mail: woodard@phys.ufl.edu
\end{flushleft}
\end{titlepage}

\section{Introduction}

Quantum gravitational back-reaction offers an attractive model of cosmology.
The idea \cite{Tsam1} is that there is no fine tuning of the cosmological
constant, $\Lambda$, or of scalar potentials. In fact there need not be any
scalars. Inflation begins in the early universe because $\Lambda$ is positive
and not unnaturally small. Inflation eventually ends due to the accumulation of
gravitational attraction between long wavelength virtual gravitons which are
pulled apart by the rapid expansion of spacetime. Inflation persists for many
e-foldings because gravity is a weak interaction, even at typical inflationary
scales, and it requires an enormous accumulation of gravitational potential to
overcome this. Since the process is infrared it can be studied reliably using
quantum general relativity, without regard to the ultraviolet problem
\cite{Tsam2}. Because the model has only a single free parameter --- $G
\Lambda$, where $G$ is Newton's constant --- it can be used to make unique and
testable predictions \cite{ATW}.

The physical mechanism of back-reaction requires quanta which are massless on
the scale of inflation but not classically conformally invariant. This rules
out competition from most ordinary matter, but it does allow an effect from
light, minimally coupled scalars. It has been suggested that significant
back-reaction can occur in scalar-driven inflation, even at one loop
\cite{MAB,ABM}. It has also been proposed that scalar self-interactions can
give a significant back-reaction at higher loops in $\Lambda$-driven inflation
\cite{Tsam3}. All these models involve fine tuning to keep the scalar light
compared with the scale of inflation, so they are probably not relevant to
phenomenology. However, scalars have the great advantage of being comparatively
simpler to study than gravitons.

With any model of back-reaction one encounters the problem of reliably
inferring its impact on the cosmological expansion rate. For a perfectly
homogeneous and isotropic geometry one would compute the expansion rate by
transforming to co-moving coordinates, reading off the scale factor, and then
taking its logarithmic time derivative. But back-reaction derives from the
gravitational response to quantum fluctuations, and these break homogeneity
and isotropy. The notion of a cosmological expansion rate must obviously have a
reasonable generalization since the current universe is not perfectly
homogeneous and isotropic, yet astronomers mean something by measuring the
Hubble constant. However, it is not so clear how to represent this observable
in terms of quantum gravitational operators.

Previous studies of back-reaction have tried to resolve this problem by
averaging over fluctuations to produce an effective geometry which is
homogeneous and isotropic. Then the cosmological expansion rate is computed
from this effective geometry in the usual way. In one method the averaging is
accomplished by taking the expectation value of the gauge fixed metric in the
presence of a state which is homogeneous and isotropic
\cite{Tsam2,Tsam3,AbWo1,AbWo2}. Then the {\it expectation value} of the metric
must be homogeneous and isotropic even though it is the average over quantum
fluctuations which are not. The other technique is to enforce homogeneity and
isotropy by spatially averaging the gauge fixed metric over a surface of
simultaneity \cite{MAB,ABM}.

Serious objections have been raised to both techniques. Unruh dislikes using
the gauge-fixed metric \cite{Unruh}, either in an expectation value or in a
spatial average. He argues that certain variations of the gauge fixing
condition change the expectation value (or spatial average) of the metric in
ways which cannot be subsumed into a coordinate transformation. Unruh therefore
maintains that even forming the expectation value (or spatial average) of the
metric into coordinate invariant quantities does not purge these quantities of
gauge dependence. He would prefer that back-reaction be studied with an
operator which is itself an invariant, before taking the expectation value. He
also disbelieves that averaging over a surface of simultaneity can be relevant
to what a local observer perceives.

A different objection has been raised by Linde. He is willing to use the gauge
fixed metric --- and both men accept the validity of quantum field theory in
determining the time evolution of the Heisenberg field operators. However,
Linde suspects that inferring back-reaction with expectation values invites a
Schrodinger Cat paradox. This is because inflationary particle production
leaves the long wavelength modes in highly squeezed states whose behavior is
essentially classical. No matter what Heisenberg operator is used to measure
the cosmological expansion rate, Linde would prefer to stochastically
\cite{LLM} sample its probability distribution rather than take its
expectation value.

The present work is an attempt to address the preceding objections. To avoid
potential problems from using the gauge fixed metric we propose to infer the
local expansion rate instead from the functional inverse of the conformal
d'Alembertian:
\begin{equation}
\label{CD}
\Box_c \equiv \frac{1}{\sqrt{-g}} \partial_{\mu} \left(\sqrt{-g} g^{\mu\nu}
\partial_{\nu} \right) - \frac16 R \; .
\end{equation}
This operator, acting on a unit ``standard candle'',
\begin{equation}
{\cal A}[g](x) \equiv \frac1{\Box_c} 1 \; ,
\end{equation}
averages over the past lightcone, as astronomers do when compiling a Hubble
diagram. In the slow roll approximation the observable gives $-\frac12 H^{-2}$
for an arbitrary homogeneous and isotropic universe. It is therefore a
reasonable candidate for measuring the local expansion rate when the universe
is not precisely homogeneous and isotropic. {\it And} it is a scalar function
of the observation point $x^{\mu}$.

Nothing can be done about the noninvariance associated with the fixed initial
value surface upon which the Heisenberg state is defined. However, invariance
under the subclass of transformations which preserve the initial value surface
can be achieved by geometrically specifying the point at which ${\cal A}[g]$ is
observed. In scalar-driven inflation this can be done by defining zero-shift
surfaces of simultaneity so that the quantum inflaton agrees with its classical
value. (Using these coordinates was Unruh's suggestion.) In more general models
one can build invariant surfaces of simultaneity using the inverse minimally
coupled d'Alembertian. The expectation value of the resulting invariant can
then be evaluated, or else its probability distribution can be sampled
stochastically.

In Section 2 we motivate the scalar and show that it has the proper
correspondence limit for exactly homogeneous and isotropic geometries. Section
3 discusses the corrections needed to geometrically specify the observation
point. In Section 4 we expand the scalar in powers of the metric fluctuations.
Section 5 concerns the retarded Green's functions which appear in this
expansion. We discuss a somewhat more complicated but considerably sharper
observable in Section 6. Our conclusions comprise Section 7. Two subsequent 
papers give the results of applying the observable to models of scalar-driven 
\cite{AbWo3} and $\Lambda$-driven \cite{AbWo4} inflation.

\section{Motivating the scalar}

Since we are interested in the effect of back-reaction on inflation it is
reasonable to consider perturbations about a background geometry which is
homogeneous, isotropic and spatially flat,
\begin{equation}
ds^2_0 = -dt^2 + e^{2 b(t)} d\vec{x} \cdot d\vec{x} = a^2(\eta) \left( -d\eta^2
+ d\vec{x} \cdot d\vec{x}\right) \; . \label{eq:background}
\end{equation}
There is general agreement that the cosmological expansion rate for this
background is $H = \dot{b} = a'/a^2$. Dots denote co-moving time derivatives
while primes represent conformal time derivatives. We normalize the initial
($t=0$ or $\eta = \eta_I$) scale factor to unity.

The full metric has the form,
\begin{equation}
g_{\mu\nu}(\eta,\vec{x}) \equiv a^2(\eta) \widetilde{g}_{\mu\nu}(\eta,\vec{x})
\equiv a^2(\eta) \left[\eta_{\mu\nu} + \kappa \psi_{\mu\nu}(\eta,\vec{x})
\right] \; ,
\end{equation}
where $\eta_{\mu\nu}$ is the spacelike Lorentz metric and $\kappa^2 \equiv 16
\pi G$ is the loop counting parameter of quantum gravity. Fluctuations reside
in the pseudo-graviton field, $\psi_{\mu\nu}(\eta,\vec{x})$, whose indices are
raised and lowered with the Lorentz metric. What we seek is a scalar functional
of the metric which provides a reasonable extrapolation for how a localized
observer would measure the cosmological expansion rate when $\psi_{\mu\nu}(
\eta,\vec{x}) \neq 0$.

It is worth explaining why the Ricci scalar is not satisfactory. $R(x)$ is
certainly scalar, and it is closely related to the Hubble constant for the case
of perfect homogeneity and isotropy,
\begin{equation}
R \longrightarrow 12 H^2 + 6 a^{-1} H' = 6 a^{-3} a^{\prime\prime} \; .
\end{equation}
However, no local curvature invariant can account for the ability of observers
to perceive the larger universe at cosmological distances by looking back along
their past light cones. Einstein's equations set the Ricci scalar to $-8 \pi G$
times the trace of the stress tensor. This actually vanishes during a phase of
radiation dominated expansion! Nor does the local value of $R(x)$ have much to 
do with what an observer can see at cosmological distances. For example, even 
``empty'' space within our solar system contains about $10$ Hydrogen atoms per 
cubic centimeter. Were we to infer the rate of cosmological expansion using 
$R(x)$ the result would correspond to a Hubble constant about a hundred times 
larger than the actual value,
\begin{equation}
8 \pi G \rho \sim 3 \cdot 10^{-30} \frac1{{\rm s}^2} \sim \left(5 \cdot 10^4
\frac{\rm km}{\rm s-Mpc} \right)^2 \; .
\end{equation}

We stress that there must be a reasonable solution to this problem because the
current universe is not precisely homogeneous and isotropic, yet human
astronomers still claim to be able to measure the Hubble constant. It is
instructive to review one of their simpler techniques. Consider light emitted
at $(\eta_1,\vec{x}_1)$ and received at $(\eta_0,\vec{x}_0)$. The observed
quantities are the redshift $z$ and the flux ${\cal F}$. If the source
luminosity ${\cal L}$ is known, these two quantities can be related under the
assumption of perfect homogeneity and isotropy. Astronomers simply define the
local Hubble constant so as to make the same relation true in the presence of
fluctuations. Then they average over many sources.

Deriving the relation between ${\cal F}$ and $z$ is a standard exercise
\cite{KT}. Assuming perfect homogeneity and isotropy the physical distance
between source and observer at time $\eta_0$ would be,
\begin{equation}
{\Delta r} \equiv a_0 \Vert \vec{x}_0-\vec{x}_1 \Vert=a_0 (\eta_0 -\eta_1) \; .
\end{equation}
The measured flux is the flat space formula, corrected for the redshifts of
energy and rate,
\begin{equation}
{\cal F} = \left({1 \over 1+ z}\right)^2 {{\cal L}\over 4\pi {\Delta r}^2} \; .
\end {equation}
One inverts this relation to solve for the product of $1+z$ times ${\Delta r}$,
which is known as the ``luminosity distance,''
\begin{equation}
d_L \equiv (1 + z) {\Delta r} = \sqrt{{\cal L} \over 4 \pi {\cal F}} \; .
\end{equation}
If both observer and source are at rest in conformal coordinates then the 
observed redshift would be,
\begin{equation}
z = {a_0 \over a(\eta_1)} - 1 \; .
\end{equation}
Its relation to ${\Delta r}$ comes from the scale factor's Taylor expansion,
\begin{equation}
a(\eta_1) = a_0 \left[1 - H_0 {\Delta r} + \frac12 (1 - q_0) H_0^2 {\Delta r}^2
+ \dots \right] \; ,
\end{equation}
where the current Hubble constant and deceleration parameter are,
\begin{equation}
H_0 \equiv {a_0' \over a_0^2} \qquad , \qquad q_0 \equiv 1 - {a_0 a_0^{\prime
\prime} \over {a_0'}^2} \; .
\end{equation}
Inverting to solve for ${\Delta r}$ gives,
\begin{equation}
H_0 {\Delta r} = z - \frac12 (1 + q_0) z^2 + \dots \; .
\end{equation}
and multiplication by $1+z$ results in the luminosity distance,
\begin{equation}
H_0 d_L = z + \frac12 (1 - q_0) z^2 + \dots \; .
\end{equation}
Plotting $z$ against $d_L$ for relatively small $z$ gives a straight line whose
inverse slope is the Hubble constant.

It is not simple to identify invariants which represent the observed quantities
$z$ and ${\cal F}$ for an arbitrary metric. If one considers the transmission 
process in terms of individual photons then the redshift could be formulated 
as follows. Let us denote the worldlines of the emitter and observer as 
functions of their respective proper times by $X^{\mu}_{\rm em}(\tau)$ and
$X^{\mu}_{\rm obs}(\tau)$. Recall that proper times are normalized to obey,
\begin{equation}
g_{\alpha \beta}\left(X(\tau)\right) \dot{X}^{\alpha}(\tau) \dot{X}^{\beta}(
\tau) = -1 \; ,
\end{equation}
where dots stand for differentiation with respect to $\tau$. Now consider a 
photon which was emitted at proper time $\tau_1$ and reaches the observer at
proper time $\tau_0$. Of course the affine parameter $\sigma$ of the photon's
worldline $X^{\mu}_{\rm ph}(\sigma)$ cannot be a proper time since the
4-velocity must be lightlike,
\begin{equation}
g_{\alpha\beta}(X_{\rm ph}(\sigma)) \dot{X}^{\alpha}_{\rm ph}(\sigma) \dot{X}^{
\beta}_{\rm ph}(\sigma) = 0 \; . \label{eq:timelike}
\end{equation}
Given any function $X^{\mu}_{\rm ph}(\sigma)$ which obeys (\ref{eq:timelike})
as it interpolates from $X^{\mu}_{\rm em}(\tau_1) = x_1^{\mu}$ to $X^{\mu}_{\rm
obs}(\tau_0) = x_0^{\mu}$, one makes a reparameterization of the affine
parameter (the new value of which we shall continue to call $\sigma$) so as to
enforce the geodesic equation,
\begin{equation}
\ddot{X}^{\mu}_{\rm ph}(\sigma) + \Gamma^{\mu}_{~\rho \sigma}\left(X_{\rm ph}(
\sigma)\right) \dot{X}^{\rho}_{\rm ph}(\sigma) \dot{X}^{\sigma}_{\rm ph}(
\sigma) = 0 \; .
\end{equation}
The redshift experienced by such a photon is given by,
\begin{equation}
1 + z = {g_{\rho\sigma}(x_1) \dot{X}^{\rho}_{\rm ph}(\sigma_1)
\dot{X}^{\sigma}_{\rm em}(\tau_1) \over g_{\mu\nu}(x_0) \dot{X}^{\mu}_{
\rm ph}(\sigma_0) \dot{X}^{\nu}_{\rm obs}(\tau_0)} \; .
\end{equation}

The flux is essentially the response, at the observer's location, to the
(presumed known) source's current density $J^{\mu}(x)$. One begins by solving
Maxwell's equations for the field strength tensor $F_{\mu\nu}(x)$,
\begin{equation}
F_{\mu\nu}^{~~;\nu} = - J_{\mu} \qquad , \qquad F_{\alpha\beta ; \gamma} +
F_{\beta\gamma ; \alpha} + F_{\gamma\alpha ; \beta} = 0 \; .
\end{equation}
These equations are invariant under local conformal rescalings,
\begin{equation}
F_{\mu\nu} \longrightarrow F_{\mu\nu} \quad , \quad J_{\mu} \longrightarrow
\Omega^2 J_{\mu} \quad , \quad g_{\mu\nu} \longrightarrow \Omega^2 g_{\mu\nu}
\; ,
\end{equation}
and they can be solved in terms of something we shall call the conformal tensor
d'Alembertian,
\begin{equation}
{_{\mu\nu}\Box}_c^{\rho\sigma} \equiv \: {_{\mu\nu}\Box}^{\rho\sigma} -
\delta^{\rho}_{~\mu} R^{\sigma}_{~\nu} - \delta^{\sigma}_{~\nu} R^{\rho}_{~\mu}
+ 2 R^{\rho ~ \sigma}_{~ \mu ~\nu} \; .
\end{equation}
The solution is,
\begin{equation}
F_{\mu\nu} = \: {_{\mu\nu}\left(\frac1{\Box_c}\right)}^{\rho\sigma} \left(-
J_{\rho ;\sigma} + J_{\sigma ; \rho}\right) \; . \label{eq:fmn}
\end{equation}
One gets the stress tensor from the field strength tensor,
\begin{equation}
T_{\mu\nu} = \left(\delta^{\alpha}_{~\mu} \delta^{\beta}_{~\nu} - \frac14
g_{\mu\nu} g^{\alpha \beta}\right) g^{\rho \sigma} F_{\alpha\rho} F_{\beta
\sigma} \; . \label{eq:tmn}
\end{equation}
The Poynting vector is obtained by contracting the observer's 4-velocity into
the electromagnetic stress tensor,
\begin{equation}
S_{\mu} = T_{\mu\nu}(x_0) \dot{X}^{\nu}(\tau_0) \; .
\end{equation}
And the measured flux is the norm of the Poynting vector in the orthogonal
projection of the observer's metric,
\begin{equation}
{\cal F}^2 = S_{\mu} S_{\nu} \left(g^{\mu\nu} + \dot{X}^{\mu}_{\rm obs}
\dot{X}^{\nu}_{\rm obs}\right) \; .
\end{equation}

Astronomers measure electromagnetic radiation because it is available to them,
but this choice of observable complicates the metric dependence of the
operators which represent their measurements. For example, the tensor character
of Maxwell's field strength is why one has to invert the tensor conformal
d'Alembertian in (\ref{eq:fmn}), rather than its simpler scalar cousin. It is
also why the response field has to be squared in (\ref{eq:tmn}). Other
complications arise from the fact that the source luminosities are not
precisely known, and that their distribution throughout space is not uniform.

The preceding complications pose important limitations on observational
astronomy but they need not restrict our choice of the operator with which to
probe the theory. For us the really essential feature is to measure the
response of some long range field to known sources distributed along the
observer's past lightcone. We can retain this feature and vastly simplify our
labor by observing a conformally coupled scalar, rather than a conformally
coupled tensor. We can achieve a further simplification by taking the source to
be a uniformly distributed monopole, rather than a sparse distribution of
dipoles of varying strength. Then a single measurement of the scalar represents
a full sky average and we can dispense with the complication of having to
tabulate two quantities ($z$ and ${\cal F}$) for each source point. We call the
scalar ${\cal A}$ and define it to obey the equation,
\begin{equation}
\Box_c {\cal A} = 1 \; , \label{eq:theeqn}
\end{equation}
where $\Box_c$ is the conformal d'Alembertian (\ref{CD}). If we define the
scalar and its first time derivative to vanish on the initial value surface the
result is just the integral of the retarded conformal Green's function,
\begin{equation}
{\cal A}[g](x) = {1 \over \Box_c} 1 \; . \label{eq:obs}
\end{equation}

It remains to show that ${\cal A}[g](x)$ has the right correspondence limit for
exact homogeneity and isotropy. In this case the conformal d'Alembertian
reduces to the form,
\begin{equation}
\Box_c \longrightarrow a^{-3} \partial^2 a \; ,
\end{equation}
where $\partial^2 \equiv \eta^{\mu\nu} \partial_{\mu} \partial_{\nu}$ is the
flat space d'Alembertian in conformal coordinates. The operator becomes even
simpler acting on spatial constants. One consequence is that (\ref{eq:theeqn})
can be solved by simple integration,
\begin{equation}
{\cal A}_0(\eta,\vec{x}) = -a^{-1}(\eta) \int_{\eta_I}^{\eta} d\eta^{\prime}
\int_{\eta_I}^{\eta'} d\eta^{\prime\prime} a^3(\eta^{\prime\prime}) =-e^{-b(t)}
\int_0^t dt' e^{-b(t')} \int_0^{t'} dt^{\prime\prime} e^{2 b(t^{\prime\prime})}
\; . \label{eq:Azero}
\end{equation}

It turns out that (\ref{eq:Azero}) can be evaluated quite generally in what is
known as {\it the slow roll approximation}. This is obeyed by all successful
models of inflation and it amounts to neglecting all higher co-moving time
derivatives of the logarithmic scale factor $b(t)$ with respect to the first,
\begin{equation}
\left\vert {d^Nb \over dt^N} \right\vert \ll \left( \dot{b} \right)^N \qquad
\forall \; N \ge 2 \; . \label{eq:SR}
\end{equation}
Most operations of ordinary calculus can be done explicitly in the slow roll
approximation. For example, the following trivial rearrangement,
\begin{equation}
e^{2b} = {d \over dt} \left({e^{2b} \over 2\dot{b}} \right) + e^{2b} {\ddot{b}
\over 2 \dot{b}^2} \; ,
\end{equation}
allows us to express the initial integrand of (\ref{eq:Azero}) as a total
derivative plus a term which is negligible in the slow roll approximation. It
would be straightforward to develop a series in slow roll corrections but the
first is generally sufficient for the inflationary setting in which we wish to
employ the new observable. With positive exponents and any significant amount
of inflation it is also possible to ignore the lower limit,
\begin{equation}
\int_0^t dt' e^{2b(t')} = \left. {e^{2b(t')} \over 2 \dot{b}(t')} \left\{ 1 +
{\ddot{b}(t') \over 2 \dot{b}^2(t')} + \dots \right\} \right\vert_0^t \approx
{e^{2 b(t)} \over 2 \dot{b}(t)} \; .
\end{equation}
We can therefore apply the slow roll approximation to Eq. (\ref{eq:Azero})
to express the observable in terms of the Hubble constant,
\begin{equation}
{\cal A}_0(\eta,\vec{x}) \approx -{1 \over 2 \dot{b}^2(t)} \; .
\end{equation}
In analogy to astronomical practice the local Hubble constant in the presence
perturbations is defined so as to preserve this relation,
\begin{equation}
{\cal A}[g](x) \equiv -{1 \over 2 H^2(x)} \; . \label{eq:slowap}
\end{equation}

Although we are chiefly interested in applying the new observable during
inflation it worth noting that the slow roll result (\ref{eq:slowap}) is valid,
up to a number of order one, for quite general geometries. For example, with
general power law expansion the logarithmic scale factor and Hubble constant
are,
\begin{equation}
b(t) = s \ln\left(1 + {H_I t \over s}\right) \qquad , \qquad \dot{b}(t) = H_I
\left(1 + \frac{H_I t}{s}\right)^{-1} \; ,
\end{equation}
where $H_I$ is the initial Hubble constant and $s$ is a constant. With this
simple time dependence we can perform the integrals in (\ref{eq:Azero})
exactly,
\begin{eqnarray}
\lefteqn{{\cal A}_{\rm power} = {- s^2 \over (s + \frac12) (s + 2)} {1 \over 2
\dot{b}^2(t)}} \nonumber \\
& & \qquad \qquad + {s^2 \over (s-1) (s+2)} {e^{-b(t)} \over H_I^2} - {s^2
\over (s-1) (s + \frac12)} {e^{-2 b(t)} \over 2 H_I \dot{b}(t)} \; .
\end{eqnarray}
The second and third terms become insignificant at late times and the first
rapidly approaches (\ref{eq:slowap}) for large $s$. Even for $s=\frac23$ the
numerical factor is $\frac17$.

\section{Fixing the observation point geometrically}

Even scalars depend upon the point at which they are observed. Part of this
dependence is physical. The Heisenberg state is specified on a particular
initial value surface and the geometrical relation of the observation point to
this initial value surface can and should affect the result. The purpose of
this section is to formulate the technology for imposing such a relation.

Of course any method of describing points amounts to fixing a gauge however,
there is an important distinction between {\it ad hoc} gauge conditions and
those which exploit some special feature of the particular system under study.
For example, if the system includes a Sun then it is geometrically meaningful
to take this star's center as the spatial origin. It is therefore necessary to
be as precise as possible about the nature of the system under study.

The dynamical variables of our system include the metric $g_{\mu\nu}(x)$ and
possibly also a scalar inflaton field $\varphi(x)$. Our goal is to compute the
expectation value of the operator ${\cal A}[g]$ (or to stochastically sample
its probability distribution) in the presence of a Heisenberg state which we
shall assume is homogeneous and isotropic. Since the only effective technique
for making such a calculation is perturbation theory we shall also assume that
the various quantum field operators are perturbations on a background which is
homogeneous and isotropic,
\begin{eqnarray}
\varphi(\eta,\vec{x}) & = & \varphi_0(\eta) + \phi(\eta,\vec{x}) \; , \\
g_{\mu\nu}(\eta,\vec{x}) & = & a^2(\eta) \left(\eta_{\mu\nu} + \kappa
\psi_{\mu\nu}(\eta,\vec{x})\right) \; .
\end{eqnarray}

Because geometrically significant gauge conditions can involve nonlocal
and nonlinear functionals of the fields, we wish to preserve the option of
carrying out the calculation in more convenient gauge. A simple technique for
accomplishing this is to define the observation point as the field-dependent
coordinate transformation $Y^{\mu}[\varphi,g](x)$ such that the transformed
scalar (if there is one) and the transformed metric,
\begin{eqnarray}
\varphi'(x) & \equiv & \varphi\left(Y(x)\right) \; , \\
g'_{\mu\nu}(x) & \equiv & {\partial Y^{\rho} \over \partial x^{\mu}} {\partial
Y^{\sigma} \over \partial x^{\nu}} g_{\rho\sigma}\left(Y(x) \right) \; ,
\end{eqnarray}
obey the geometrically significant gauge conditions. Then one can evaluate
${\cal A}[g](Y(x))$ in any gauge and the result will be the same.

The existence of a fixed initial value surface (at $\eta=\eta_I$) suggests that
$Y^{\mu}$ should be expressed as the composition of a temporal transformation
$\eta \longrightarrow \tau(\eta,{\vec x})$ followed by a purely spatial
transformation $x^i \longrightarrow \chi^i(\eta,{\vec x})$. Surfaces of
simultaneity are defined by the condition $\tau(\eta,\vec{x}) ={\rm constant}$,
while $\chi^i(\eta,\vec{x})$ traces out ``the same'' space point on the
foliation of these surfaces. The full transformation would be,
\begin{equation}
Y^0(\eta,\vec{x}) = \tau(\eta,\vec{x}) \quad , \quad Y^i(\eta,\vec{x}) = 
\chi^i\left(\tau(\eta,\vec{x}),\vec{x}\right) \; . \label{eq:trans}
\end{equation}

The problem's homogeneity and isotropy implies that all space points are
physically equivalent and we may as well use orthogonal projection to define
``the same'' space point. This amounts to the condition $g^{\prime}_{0i} = 0$
and hence,
\begin{equation}
0 = {\partial \chi^k \over \partial x^i} g_{0k}(\eta,{\vec \chi}) + {\partial
\chi^j \over \partial \eta} {\partial \chi^k \over \partial x^i} g_{jk}(\eta,
\vec{\chi}) \; .
\end{equation}
The relation can be simplified by multiplying with the inverse Jacobian,
\begin{equation}
g_{0j}(\eta,\vec{\chi}) + {\partial \chi^i \over \partial \eta} g_{ij}(\eta,
\vec{\chi}) = 0 \; .
\end{equation}
Whereupon multiplication by the inverse 3-metric results in the following first
order (but nonlinear) differential equation,
\begin{eqnarray}
{\partial \chi^i \over \partial \eta} & = & -\left(g^{-1}\right)^{ij} g_{0j}(
\eta,\vec{\chi}) \; , \\
& = & -\kappa \psi_{0i}(\eta,\vec{\chi}) + \kappa^2 \left(\psi_{0j} \psi_{ji}
\right)(\eta,\vec{\chi}) - \kappa^3 \left(\psi_{0k} \psi_{kj} \psi_{ji}\right)(
\eta,\vec{\chi}) + \dots \quad
\end{eqnarray}
Making the obvious choice of initial condition gives an integral equation whose
iteration to any order is straightforward,
\begin{eqnarray}
\chi^i(\eta,\vec{x}) & = & x^i - \int_{\eta_I}^{\eta} d\sigma \left(g^{-1}
\right)^{ij} g_{0j}(\sigma,\vec{\chi}) \; , \label{chi} \\
& = & x^i - \kappa \int_{\eta_I}^{\eta} d\sigma \psi_{0i}(\sigma,\vec{x}) +
\kappa^2 \int_{\eta_I}^{\eta} d\sigma \left(\psi_{0j} \psi_{ji}\right)(\sigma,
\vec{x}) \nonumber \\
& & \qquad + \kappa^2 \int_{\eta_I}^{\eta} d\sigma \,
\psi_{0i,j}(\sigma,\vec{x}) \int_{\eta_I}^{\sigma} d\rho \,
\psi_{0j}(\rho,\vec{x}) + O(\kappa^3) \; .
\end{eqnarray}

Defining surfaces of simultaneity is less subtle for $\Lambda$-driven inflation
than for its scalar-driven cousin. Without back-reaction the cosmological
expansion rate is constant in $\Lambda$-driven inflation, so the effect is
certainly real if one sees progressive slowing under any timelike foliation. In
the scalar-driven case there is already slowing as the background scalar rolls
down its potential so one must be careful to compare the expansion rate with
and without back-reaction at the same physical time.

Since the value of the scalar determines the expansion rate without
back-reaction it seems reasonable to define surfaces of simultaneity so that
the full inflaton field agrees with its background value,
\begin{equation}
\varphi\left(\tau(\eta,\vec{x}),\vec{x}\right) \equiv \varphi_0(\eta) \; . 
\label{scalar}
\end{equation}
This can be solved perturbatively by first writing,
\begin{equation}
\tau(\eta,\vec{x}) = \eta + {\delta \tau}(\eta,\vec{x}) \; ,
\end{equation}
and Taylor expanding,
\begin{equation}
\sum_{n=1}^{\infty} {\varphi_0^{(n)}(\eta) \over n!} \left({\delta \tau}(\eta,
\vec{x})\right)^n = -\sum_{n=0}^{\infty} {\phi^{(n)}(\eta,\vec{x}) \over n!}
\left({\delta \tau}(\eta,\vec{x})\right)^n \; .
\end{equation}
Inverting results in an expansion for ${\delta \tau}$ in powers of the quantum
scalar $\phi$ and its derivatives (all evaluated at $(\eta,\vec{x})$),
\begin{equation}
{\delta \tau} = - {\phi \over \varphi'} + {\phi \phi' \over \varphi_0^{\prime
2}} - {\varphi_0^{\prime\prime} \phi^2 \over 2 \varphi_0^{\prime 3}} +
O(\phi^3) \; .
\end{equation}

$\Lambda$-driven inflation can be included within the same scheme by employing
a scalar functional of the metric with monotonic time dependence in place of 
the dynamical scalar. Perhaps the simplest of these ``clock functions'' makes 
use of the inverse minimally coupled d'Alembertian acting on the Ricci scalar,
\begin{equation}
\label{FOP}
{\cal N}[g](x) \equiv - \frac{1}{4\Box} R \; .
\end{equation}
We define surfaces of simultaneity so as to make the clock agree with its 
background value, just as relation (\ref{scalar}) does for scalar-driven
inflation,
\begin{equation}
{\cal N}\left(\tau(\eta,\vec{x}),\vec{x}\right) \equiv {\cal N}_0(\eta) \; .
\end{equation}
Fig.~1 depicts the resulting foliation.

\begin{figure}
\centerline{\epsfxsize=0.6\textwidth\epsffile{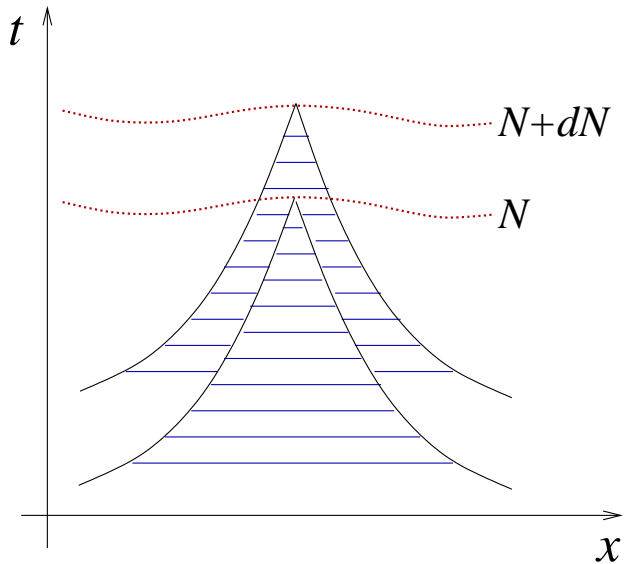}}
\caption{Invariant procedure to fix the observation point.}
\end{figure}

To see that ${\cal N}[g](x)$ is a good clock in perturbation theory note that 
${\cal N}_0(\eta) \approx \ln(a(\eta))$ in the slow roll approximation. This 
follows because the minimally coupled d'Alembertian takes the following form in
a homogeneous, isotropic and spatially flat geometry,
\begin{equation}
\Box \equiv {1 \over \sqrt{-g}} \partial_{\mu} \left(\sqrt{-g} g^{\mu\nu}
\partial_{\nu}\right) \longrightarrow a^{-4} \partial_{\mu} \left( a^2 \eta^{
\mu\nu} \partial_{\nu}\right) \; .
\end{equation}
When acting on functions of only time the inverse of $\Box$ reduces to,
\begin{equation}
\frac1{\Box} \longrightarrow -\int_{\eta_I}^{\eta} d\eta' a^{-2}(\eta') \int_{
\eta_I}^{\eta'} d\eta^{\prime\prime} a^4(\eta^{\prime\prime}) = - \int_0^t dt'
e^{-3 b(t')} \int_0^{t'} dt^{\prime\prime} e^{3 b(t^{\prime\prime})} \; .
\end{equation}
With no perturbations we therefore have,
\begin{equation}
{\cal N}_0(\eta) = \int_0^t dt' e^{-3 b(t')} \int_0^{t'} dt^{\prime\prime} 
e^{3 b(t^{\prime\prime})} \left(3 \dot{b}^2(t^{\prime\prime}) + \frac32 
\ddot{b}(t^{\prime\prime})\right) \; .
\end{equation}
Making the slow roll approximation gives,
\begin{eqnarray}
{\cal N}_0(\eta) & \approx & \int_0^t dt' e^{-3 b(t')} \int_0^{t'} dt^{\prime
\prime} {d \over dt^{\prime\prime}} \left(\dot{b}(t^{\prime\prime}) e^{3 b(t^{
\prime \prime})}\right) \; , \\
& \approx & \int_0^t dt' \dot{b}(t') = b(t) \; .
\end{eqnarray}

Before closing the section we should comment that there is no problem, in
perturbation theory, about evaluating an operator such as ${\cal A}[g](x)$ at
a point such as $Y^{\mu}[\varphi,g](x)$ which is itself an operator. One merely
expands in powers of the perturbatively small quantity $Y^{\mu}(x) - x^{\mu}$.
Only a finite number of terms need be included to reach any fixed order in
perturbation theory. Note that the various operator products should be
time-ordered. This is because the functional integral of the $\comp$-number
functional ${\cal A}[g]\left(Y[\varphi,g](x)\right)$ is manifestly invariant as
it is, and gives the expectation value of the time-ordered product of the
corresponding operator.

\section{Pseudo-graviton expansion}

The purpose of this section is to expand ${\cal A}[g](x)$ in powers of the
pseudo-graviton field $\psi_{\mu\nu}(\eta,\vec{x})$. This is most easily
accomplished by first expressing $\Box_c$ in terms of the conformally rescaled
metric,
\begin{equation}
\widetilde{g}_{\mu\nu}(\eta,\vec{x}) \equiv a^{-2}(\eta) g_{\mu\nu}(\eta,\vec{
x}) = \eta_{\mu\nu} + \kappa \psi_{\mu\nu}(\eta,\vec{x}) \; .
\end{equation}
We now write ${\Box}_c = a^{-3} {\cal D} a$, where ${\cal D}$ and its expansion
are,
\begin{equation}
{\cal D} \equiv {1 \over \sqrt{-\widetilde{g}}} \partial_{\mu} \left(\sqrt{-
\widetilde{g}} \: \widetilde{g}^{\mu\nu} \partial_{\nu}\right) - \frac16
\widetilde{
R} = \partial^2 + \kappa {\cal D}_1 + \kappa^2 {\cal D}_2 + \dots \; .
\end{equation}
The first two operators in the expansion are,
\begin{eqnarray}
{\cal D}_1 &=& - \psi^{\mu\nu} \partial_\mu \partial_\nu + \left(-\psi^{\mu
\alpha}_{~~, \alpha} + \frac12 \psi^{,\mu} \right) \partial_{\mu} - \frac16
\left( \psi^{\rho\sigma}_{~~ , \rho\sigma} - \psi^{,\rho}_{~~\rho}\right) \; ,
\label{eq:D1} \\
{\cal D}_2 &=& \psi^{\mu\alpha} \psi^{\nu}_{\; \alpha} \partial_{\mu}
\partial_{\nu} + \left( (\psi^{\alpha\beta} \psi^{\mu}_{~\alpha})_{, \beta} -
\frac12 \psi^{\alpha\beta,\mu} \psi_{\alpha\beta} + \frac12 \psi^{\alpha\mu}
\psi_{,\alpha}\right) \partial_{\mu} -\frac16 \widetilde{R}_2 . \qquad
\label{D2}
\end{eqnarray}
We remind the reader that pseudo-graviton indices are raised and lowered by
the Lorentz metric, $\eta_{\mu\nu}$. Other notational points are that the
trace of the pseudo-graviton field is $\psi \equiv \psi^{\rho}_{~ \rho}$ and
that the second order, conformally rescaled Ricci scalar is,
\begin{eqnarray}
\lefteqn{\widetilde{R}_2 \equiv \psi^{\alpha\beta} \left(\psi^{~~~ ,\gamma}_{
\alpha\beta ~~ \gamma} + \psi_{,\alpha\beta} - 2 \psi^{\gamma}_{~ \alpha ,
\beta \gamma} \right) } \nonumber \\
& & + \frac34 \psi^{\alpha\beta,\gamma} \psi_{\alpha\beta,\gamma} - \frac12
\psi^{\alpha \beta ,\gamma} \psi_{\gamma \beta,\alpha} - \psi^{\alpha \beta}_{
~~~ ,\beta} \psi^{\gamma}_{~ \alpha , \gamma} +\psi^{\alpha \beta}_{~~~ ,\beta}
\psi_{, \alpha} - \frac14 \psi^{,\alpha} \psi_{, \alpha} \; . \quad
\end{eqnarray}

The next step is to factor $\partial^2$ out of ${\cal D}$,
\begin{equation}
{\cal D} = \partial^2 \left(1 + \frac1{\partial^2} \kappa {\cal D}_1 + \frac1{
\partial^2} \kappa^2 {\cal D}_2 + O(\kappa^3)\right) \; .
\end{equation}
Inverting ${\cal D}$ is now straightforward,
\begin{equation}
\frac1{\cal D} = \frac{1}{\partial^2} - \frac{1}{\partial^2} \kappa {\cal
D}_1 \frac{1}{\partial^2} + \frac1{\partial^2} \kappa {\cal D}_1 \frac1{
\partial^2} \kappa {\cal D}_1 \frac{1}{\partial^2} - \frac{1}{\partial^2}
\kappa^2 {\cal D}_2 \frac{1}{\partial^2} + O(\kappa^3) \; .
\end{equation}
All this implies the following expansion for the scalar observable,
\begin{equation}
{\cal A}[g] = a^{-1} \frac1{\cal D} a^3 = {\cal A}_0 + \kappa {\cal A}_1 +
\kappa^2 {\cal A}_2 + O(\kappa^3) \; . \label{A_exp}
\end{equation}
${\cal A}_0$ was worked out at the end of Section 2. The next two terms are,
\begin{eqnarray}
{\cal A}_1 & \equiv & - a^{-1} \frac1{\partial^2} {\cal D}_1 \frac1{\partial^2}
a^3 \; , \\
{\cal A}_2 & \equiv & - a^{-1} \frac1{\partial^2} {\cal D}_2 \frac1{\partial^2}
a^3 + a^{-1} \frac1{\partial^2} {\cal D}_1 \frac1{\partial^2} {\cal D}_1
\frac1{\partial^2} a^3 \; .
\end{eqnarray}

It should be noted that the pseudo-graviton field is not free, nor are all of
its components dynamical. The next step after this would be to expand $\psi_{
\mu\nu}(\eta,\vec{x})$ in terms of the fundamental dynamical degrees of
freedom, whatever they happen to be.  This obviously depends upon selecting a
particular model and must be postponed until this has been done
\cite{AbWo3,AbWo4}.

\section{Retarded Green's functions}

The pseudo-graviton expansion of the previous section results in a series of
terms which involve the inverse differential operator $1/\partial^2$. The
purpose of this section is to precisely define the action of this operator.
We also apply the slow roll approximation.

The first task is easily accomplished. The retarded Green's function for the
operator $\partial^2$ is well known,
\begin{equation}
G(x;x') = -\frac{\theta(\Delta \eta)}{ 4\pi {\Delta x}} \delta (\Delta \eta
- \Delta x ) \; ,
\end{equation}
where $\Delta \eta \equiv \eta - \eta'$ and $\Delta x \equiv \Vert\vec{x} -
\vec{x}'\Vert$. Since the initial value surface is at $\eta = \eta_I$ we define
the result of acting $1/\partial^2$ on an arbitrary function $f(\eta,\vec{x})$
as,
\begin{eqnarray}
\left[\frac{1}{\partial^2} f\right](\eta,\vec{x}) & \equiv & -\int_{\eta_I}^{
\eta} d\eta' \int d^3x' {\delta({\Delta \eta} - {\Delta x}) \over 4\pi {\Delta
x}} f(\eta',\vec{x}') \; , \\
& = & -\int_{\eta_I}^{\eta} d\eta' {\Delta \eta} \int {d^2\widehat{n} \over
4\pi} f\left(\eta',\vec{x} + {\Delta \eta} \widehat{n}\right) \; .
\end{eqnarray}

When the function depends only upon time we can reach a form similar to that of
Section 2. Making the substitution $f(\eta,\vec{x}) \longrightarrow F(\eta)$
gives,
\begin{eqnarray}
\left[\frac1{\partial^2} F\right](\eta,\vec{x}) & = & -\int_{\eta_I}^{\eta}
d\eta' {\Delta \eta} F(\eta') \; , \\
& = & -\int_{\eta_I}^{\eta} d\eta' (\eta - \eta') {d \over d\eta'} \int_{
\eta_I}^{\eta'} d\eta^{\prime\prime} F(\eta^{\prime\prime}) \; , \\
& = & -\int_{\eta_I}^{\eta} d\eta' \int_{\eta_I}^{\eta'} d\eta^{\prime\prime}
F(\eta^{\prime\prime}) \; .
\end{eqnarray}
An important example is provided by the rightmost term for each of the ${\cal
A}_n$'s --- $\partial^{-2} a^3$. We can explicitly evaluate these terms by
making use of the slow roll approximation,
\begin{equation}
\left[\frac1{\partial^2} a^3\right](\eta,\vec{x}) = -\int_0^t dt' e^{-b(t')}
\int_0^{t'} dt^{\prime\prime} e^{2 b(t^{\prime\prime})} \approx {-e^{b(t)}
\over 2 \dot{b}^2(t)} \; .
\end{equation}
The fact that this depends only upon $\eta$ allows one to simplify expressions
in which derivatives act upon it. For example, the contribution from the first
term in (\ref{eq:D1}) is,
\begin{eqnarray}
\lefteqn{a^{-1} \frac1{\partial^2} \kappa \psi^{\mu\nu} \partial_{\mu}
\partial_{\nu} \frac{1}{\partial^2} a^3 = - a^{-1} \frac1{\partial^2} \kappa
\psi_{00} a^3} \; , \\
& = & a^{-1}(\eta) \int_{\eta_I}^{\eta} d\eta' a^3(\eta') {\Delta \eta} \int
{d^2\widehat{n} \over 4\pi} \kappa \psi_{00}\left(\eta',\vec{x} + {\Delta \eta}
\widehat{n}\right) \; .
\end{eqnarray}

Further progress requires using the slow roll approximation. It turns out, as
one evaluates the various factors of $1/\partial^2$ from left to right, that
the various integrands upon which they act are always dominated by the
universal initial factor of $a^3$. In typical gauges the pseudo-graviton field
can grow at most like powers of $\ln(a)$. Although derivatives can sometimes
result in a net loss of powers of the scale factor, they can never add such
powers. Further, whenever even a single power of $a$ is lost the contribution
which finally results to ${\cal A}[g](x)$ is exponentially suppressed and hence
irrelevant. It therefore suffices to consider terms of the form $\partial^{-2}
(a^3 f)$ for functions $f(\eta,\vec{x})$ which grow less rapidly that
$a(\eta)$,
\begin{equation}
\left[\frac1{\partial^2} a^3 f\right](\eta,\vec{x}) = -\int_0^t dt' e^{2 b(t')}
{\Delta \eta} \int {d^2\widehat{n} \over 4\pi} f\left(\eta',\vec{x} + {\Delta
\eta} \widehat{n}\right) \; . \label{eq:generic}
\end{equation}
Note that ${\Delta \eta}$ and $\eta'$ are the following functions of $t$ and
$t'$,
\begin{equation}
{\Delta \eta} = \int_{t'}^t dt^{\prime\prime} e^{-b(t^{\prime\prime})} \qquad ,
\qquad \eta' = \eta_I + \int_0^{t'} dt^{\prime\prime} e^{-b(t^{\prime\prime})}
\; .
\end{equation}

Because ${\Delta \eta}$ vanishes at $t'=t$ a single partial integration fails
to extract the leading order term in the slow roll approximation,
\begin{equation}
\int_0^t dt' e^{2 b(t')} {\Delta \eta} f\left(\eta',\vec{x} + {\Delta \eta}
\widehat{n}\right) = \left. e^{2b} {{\Delta \eta} f \over 2 \dot{b}}
\right\vert_0^t - \int_0^t dt' e^{2b} {d \over dt'} \left[{{\Delta \eta} f
\over 2 \dot{b}}\right] \; .
\end{equation}
The surface term vanishes at the upper limit and is exponentially suppressed
at the lower limit. The remaining integrand is,
\begin{equation}
e^{2b} {d \over dt'} \left({{\Delta \eta} f \over 2 \dot{b}}\right) = -
{\ddot{b} \over 2 \dot{b}^2} e^{2b} {\Delta \eta} f + {e^{b} \over 2 \dot{b}}
\left\{ {\mbox{} \over \mbox{}} -f + {\Delta \eta} \partial_0 f - {\Delta \eta}
\widehat{n} \cdot \vec{\nabla} f\right\} \; .
\end{equation}
The first term on the right is the original integrand times a term which is
negligible in the slow roll approximation. Of the remaining terms only the one
without the factor of ${\Delta \eta}$ survives at the upper limit after another
partial integration. Since each additional partial integration produces either
a factor of $\ddot{b}/\dot{b}^2$ or of $e^{-b}$ the slow roll approximation of
(\ref{eq:generic}) is,
\begin{eqnarray}
\left[\frac1{\partial^2} a^3 f\right](\eta,\vec{x}) & \approx & \int_0^t dt'
{e^{b(t')} \over 2 \dot{b}(t')} \int {d^2\widehat{n} \over 4\pi} f\left(\eta',
\vec{x} + {\Delta \eta} \widehat{n}\right) \; , \\
& \approx & - {a(\eta) \over 2 H^2(\eta)} f(\eta,\vec{x}) \; . \label{eq:simp}
\end{eqnarray}

The vast simplification inherent in (\ref{eq:simp}) derives from the fact that
the response of a conformally coupled scalar at $\eta = \eta_0$ to that part of
the source at $\eta = \eta_1$ redshifts as $a(\eta_0)/a(\eta_1)$. Hence only
the most recent sources contribute effectively. One can recapture the slow roll
approximation much more simply by rewriting the differential equation
(\ref{eq:theeqn}) which defines ${\cal A}[g](x)$,
\begin{equation}
{\cal A} = -{6 \over R} \left(1 - {1 \over \sqrt{-g}} \partial_{\mu} \sqrt{-g}
g^{\mu\nu} \partial_{\nu} {\cal A}\right) \; .
\end{equation}
Since the Ricci scalar is almost constant during inflation the leading order
slow roll term --- to {\it all} orders in the pseudo-graviton expansion --- is
contained in the reduction, ${\cal A}[g](x) \longrightarrow -6/R(x)$.

\section{A better observable}

Simple expressions are nice but the results of the previous section are too 
much of a good thing. To leading order in the slow roll approximation our new
observable has turned out to be nothing more than $-6$ over the Ricci scalar!
The next order terms disrupt this correspondence but still give expressions
which are local in the observation point. We criticized this sort of locality 
in section 2. A reasonable measure of the cosmological expansion rate should 
not be dominated by local fluctuations. Of course small fluctuations are, by 
definition, sub-dominant to the background, so one can still use ${\cal A}[g](
x)$ to measure back-reaction during the first stages of inflation. However, the
defect of too much locality has a relatively simple fix which we shall present
in this section. We shall also refine the observable so that it gives the 
Hubble constant exactly for any homogeneous and isotropic geometry, without 
recourse to the slow roll approximation.

The effective locality of ${\cal A}[g](x)$ derives from the fact that conformal 
scalars redshift like the inverse scale factor. During inflation the scale
factor grows so rapidly that only the most recent sources matter much. A 
straightforward way of avoiding this is by breaking conformal invariance. 
Suppose we measure a minimally coupled scalar ${\cal B}[g](x)$, whose value and
whose first derivative vanish on the initial value surface, and which is driven
by a source $S(x)$,
\begin{equation}
\Box {\cal B}(x) = S(x) \; .
\end{equation}
For a homogeneous and isotropic geometry (\ref{eq:background}) the slow roll
approximation results in an almost uniformly weighted average over comoving 
time,
\begin{eqnarray}
\frac1{\Box} S & \longrightarrow & - \int_0^t dt' e^{3 b(t')} \int_0^t dt^{
\prime\prime} e^{3 b(t^{\prime\prime})} S(t^{\prime\prime}) \\
& \approx & - \int_0^t dt' {S(t') \over 3 \dot{b}(t')} \; .
\end{eqnarray}

It remains to identify a suitable source. Note that for a homogeneous and
isotropic geometry (\ref{eq:background}) the nonzero components of the
Ricci tensor are,
\begin{equation}
R_{00} \longrightarrow -3 \ddot{b} -3 \dot{b}^2 \qquad , \qquad R_{ij}
\longrightarrow (\ddot{b} + 3 \dot{b}^2) g_{ij} \; .
\end{equation}
The trace of the spatial part has the curious property of giving a total
derivative when multiplied by $e^{3 b}$,
\begin{equation}
3 \left(\ddot{b}(t) + 3 \dot{b}^2(t)\right) e^{3b(t)} = {d \over dt} \left(3
\dot{b}(t) e^{2 b(t)} \right) \; .
\end{equation}
If the source $S(x)$ reduces to minus one third times this spatial trace, for a 
homogeneous and isotropic geometry, then the minimally coupled scalar will 
reduce to the logarithmic scale factor exactly,
\begin{equation}
{\cal B}[g](x) \longrightarrow \int_0^t dt' e^{3 b(t')} \int_0^{t'} dt^{\prime
\prime} e^{3b(t^{\prime\prime})} \left(\ddot{b}(t^{\prime\prime}) + 3 
\dot{b}^2(t^{\prime\prime})\right) = b(t) \; .
\end{equation}
We can then obtain the Hubble constant by differentiation with respect to
comoving time.

The object we have just described is not quite a scalar because the spatial
trace of the Ricci tensor is not. However, we can give the latter an invariant 
formulation by exploiting the technology of section 3 to define it in a special
coordinate system which reduces to conformal coordinates for a homogeneous and 
isotropic geometry. With the transformation $Y^{\mu}[g](x)$ we can define the 
spatial components of the metric and the Ricci tensor,
\begin{equation}
g'_{ij}(x) \equiv {\partial Y^{\rho} \over \partial x^i} {\partial Y^{\sigma}
\over \partial x^j} g_{\rho \sigma}\left(Y(x)\right) \quad , \quad
R'_{ij}(x) \equiv {\partial Y^{\rho} \over \partial x^i} {\partial Y^{\sigma}
\over \partial x^j} R_{\rho \sigma}\left(Y(x)\right) \; .
\end{equation}
The 3-curvature is just the inverse of the first contracted into the second,
\begin{equation}
{\cal R}[g](x) \equiv \left(g^{\prime -1}\right)^{ij} R'_{ij}(x) \; ,
\end{equation}
and the minimally coupled scalar is,
\begin{equation}
{\cal B}[g](x) = \frac1{\Box} \left(-\frac13 {\cal R}\right) \; .
\end{equation}

Since conformal coordinates have zero shift we obviously wish to use the same
spatial transformation (\ref{chi}) as in section 3. The temporal transformation
requires a scalar clock function, the most uniformly applicable choice of which
is ${\cal V}[g](x)$, the invariant volume of the past lightcone as seen from 
the point $x^{\mu}$ back to the inital value surface. Note that it must be a
good clock generally, not just in perturbation theory, because the volume of
the past lightcone increases monotonically under any timelike foliation. 

We define ${\cal V}[g](x)$ as the invariant integral over all points which are 
connected to $x^{\prime \mu}$ by any future-directed, non-spacelike 
path.\footnote{The path need not be a geodesic, nor does it have to be the sole
path which connects $x^{\prime \mu}$ and $x^{\mu}$.} For a homogeneous and 
isotropic geometry this reduces to a single integral,
\begin{equation}
{\cal V}[g](x) \longrightarrow {\cal V}_0(\eta) \equiv \frac43 \pi \int_{\eta_I
}^{\eta} d\eta' \Omega^4(\eta') (\eta - \eta')^3 \; .
\end{equation}
As in section 3, we define surfaces of simultaneity to make this relation
persist in the presence of perturbations,
\begin{equation}
{\cal V}\left(\tau(\eta,\vec{x}),\vec{x}\right) \equiv {\cal V}_0(\eta) \; .
\end{equation}

We define the general conformal factor as the square root of the $00$ component
of the metric in these coordinates,
\begin{eqnarray}
\Omega[g](x) & \equiv & \sqrt{g'_{00}(x)} \; , \\
& = & {\partial \tau \over \partial \eta} \left[{\mbox{} \over \mbox{}} 
g_{00}(Y(x)) - \left(g^{-1}\right)^{ij} g_{0i} g_{0j}(Y(x))\right]^{\frac12} 
\; ,
\end{eqnarray}
Since the coordinates have zero shift, $g'_{0i} = 0$, and it follows that
$\Omega$ is precisely the factor needed to scale from conformal to co-moving 
time. One possible definition for the Hubble constant is therefore the 
co-moving time derivative of the scalar ${\cal B}$ evaluated in these 
coordinates,
\begin{equation}
{\cal H}_1[g](x) \equiv \Omega^{-1}[g](x) {\partial \over \partial \eta}
{\cal B}[g]\left(Y[g](x)\right) \; .
\end{equation}

Of course one might equally well base the observable on $\Omega[g](x)$ now that
we have it,
\begin{equation}
{\cal H}_2[g](x) \equiv \Omega^{-1}[g](x) {\partial \over \partial \eta}
\ln\left({\mbox{} \over \mbox{}} \Omega[g](x)\right) \; .
\end{equation}
One might instead employ the third root of the determinant of $g'_{ij}(x)$. All
of these are plausible measures for the cosmological expansion rate, all reduce
exactly to the Hubble constant for homogeneous and isotropic geometries, and
we anticipate that all will give the same result as regards the existence or
non-existence of a significant back-reaction. We emphasize this multiplicity of
plausible observables is as it should be because a similar situation exists in
the many different methods by which astronomers attempt to measure Hubble
constant.

\section{Discussion}

Reliably quantifying the effect of back-reaction on inflation poses a
frustrating paradox. The possibility of an effect derives from fluctuations in
homogeneity and isotropy, but these call into question precisely what is meant
by the rate of cosmological expansion. Previous work has attempted to resolve
the issue by averaging the gauge fixed metric, either over a surface of
simultaneity \cite{MAB,ABM} or over the range of quantum fluctuations in a
homogeneous and isotropic state \cite{Tsam2,Tsam3,AbWo1,AbWo2}. It has been
objected that neither technique is manifestly invariant, and also that the
former procedure involves superposing data unavailable to a local observer on
the surface of simultaneity \cite{Unruh}.

In section 2 we argued that both problems can be avoided by measuring the
response of a noninteracting, conformally coupled scalar to a constant source.
The scalar ${\cal A}[g](x)$ is a nonlocal functional of the metric which is
obtained by superposing over the past light cone of $x^{\mu}$, just as
astronomers do in measuring the Hubble constant. Its phenomenological
interpretation also follows the standard practice in astronomy: we define the
locally observed rate of cosmological expansion to bear the same relation ---
equation (\ref{eq:slowap}) --- to ${\cal A}[g](x)$ for a general metric as it
does for a homogeneous and isotropic one. Of course the result will be a little
different at different locations, just as we must expect the Hubble constant
measured by human astronomers to disagree slightly with the value obtained from
the different field of view available to their opposite numbers in the Coma
Cluster. But nearby observers will tend to agree because their past lightcones
largely overlap.

It should be noted that we are not adding a conformal scalar to whatever
model of inflation is being probed. The observable is only a functional of the
metric used to pose invariant questions about the expansion rate; it does not
change the dynamics of the model. Even if one insists that the scalar
represents a sort of measuring device whose effect must be included, the
strength of the constant source can still be adjusted so as to make this effect
negligible. The required constant would simply appear on the right hand sides
of both equations (\ref{eq:obs}) and (\ref{eq:slowap}),
\begin{equation}
{\cal A}[g](x) \longrightarrow \frac1{\Box_c} K \equiv {-K \over 2 H^2(x)} \; ,
\end{equation}
so that the magnitude of the scalar could be made arbitrarily small without
affecting our determination of the local Hubble constant.

One can either compute the expectation value of ${\cal A}[g](x)$ --- or else
sto\-chastically sample its probability distribution --- in the presence of a
Heisenberg state which we assume to be homogeneous and isotropic. The passage
from a scalar to an invariant can be achieved by geometrically fixing the
observation point relative to the initial value surface on which the Heisenberg
state is defined. Because all points on the initial value surface are
physically equivalent the problem reduces to orthogonally projecting between
geometrically specified surfaces of simultaneity. Two definitions for such
surfaces were presented in Section 3, along with perturbative expansions for
the field dependent observation point $Y^{\mu}[\varphi,g](x)$ which can be used
to fix the observation point when working an arbitrary gauge.

Sections 4 and 5 developed the general machinery necessary to evaluate the new
observable perturbatively. We emphasize that these computations are imminently
doable in the slow roll approximation. It remains to apply the technology to
simple models of scalar-driven \cite{AbWo3} and $\Lambda$-driven \cite{AbWo4}
inflation.

An embarrassing postscript to these labors is that the slow roll approximation
purges ${\cal A}[g](x)$ of its nonlocality. In fact it reduces to $-6/R(x)$,
which we initially rejected as being dominated by local fluctuations! There is
actually no obstacle to making use of ${\cal A}[g](x)$ in perturbation theory 
because small fluctuations are, by definition, sub-dominant to the homogeneous 
background. However, one would still prefer an observable which represents a 
more evenly weighted average over the past lightcone. Several alternatives are 
discussed in section 6. We have taken the additional trouble to construct them
to reduce exactly to the Hubble constant for homogeneous and isotropic 
geometries, without recourse to the slow roll approximation.

\vskip 1cm
\centerline{\bf Acknowledgments}

It is a pleasure to acknowledge stimulating and informative conversations with
R. Bond, R. Brandenberger, A. Guth, L. Kofman, A. Linde, D. Lyth, V.
Mukhanov, E. Stewart and W. Unruh. We are also grateful to the University of
Crete and to the Aspen Center for Physics for their hospitality during
portions of this project. This work was partially supported by DOE contract
DE-FG02-97ER\-41029, by NSF grant 94092715 and by the Institute for
Fundamental Theory.

\end{document}